\documentclass[manuscript]{aastex}
\usepackage{longtable}

\newcommand{\eqmy}{Eq.~\ref{rho2} }
\newcommand{\eqsmo}{Eq.~\ref{rho1} }

\shorttitle{Using $\rho_\star$ to Evaluate Transit Candidates}
\shortauthors{Tingley, Bonomo \& Deeg}

\begin{document}

\title{Using Stellar Densities to Evaluate Transiting Exoplanetary
       Candidates}

\author{B. Tingley}

\affil{Insituto de Astrof\'{i}sica de Canarias, E38205 - La Laguna
  (Tenerife), España}
\affil{Dpto. de Astrof\'isica, Universidad de La Laguna, 38206 La
  Laguna, Tenerife, Spain}
\email{btingley@iac.es}

\author{A.~S. Bonomo}
\affil{ Observatoire Astronomique de Marseille-Provence,
 13388 Marseille cedex 13, France}

\and

\author{H.~J. Deeg}
\affil{Insituto de Astrof\'{i}sica de Canarias,
    E38205 - La Laguna (Tenerife), España}
\affil{Dpto. de Astrof\'isica, Universidad de La Laguna, 38206 La
  Laguna, Tenerife, Spain}

\begin{abstract}
  One of the persistent complications in searches for transiting
  exoplanets is the low percentage of the detected candidates that
  ultimately prove to be planets, which significantly increases the
  load on the telescopes used for the follow-up observations to
  confirm or reject candidates. Several attempts have been made at
  creating techniques that can pare down candidate lists without the
  need of additional observations. Some of these techniques involve a
  detailed analysis of light curve characteristics; others estimate
  the stellar density or some proxy thereof. In this paper, we extend
  upon this second approach, exploring the use of
  independently-calculated stellar densities to identify the most
  promising transiting exoplanet candidates. We use a set of CoRoT
  candidates and the set of known transiting exoplanets to examine the
  potential of this approach. In particular, we note the possibilities
  inherent in the high-precision photometry from space missions, which
  can detect stellar asteroseismic pulsations from which accurate
  stellar densities can be extracted without additional observations.
\end{abstract}


\section{Introduction}

One of the principle goals of exoplanet searches using transits is the
discovery of temperate terrestrial exoplanets. Their detection
presents a tremendous challenge -- particularly for the radial
velocity observations that are typically used to confirm a transiting
exoplanet candidate. Even the confirmation of transiting Hot Jupiters
can confound observers, consuming and ultimately wasting telescope
resources. Transit surverys typically identify 10-20 candidates for
each planet (STARE, WASP \citep{poll2006}, OGLE
\citep{uda2002a,uda2002b,uda2002c,uda2003}, HAT \citep{bak2004}),
although, in the case of CoRoT, the ratio of top priority candidates
from that mission is closer to one-third or even one-half
\citep{cab2009}. This demonstrates the success of a decade-long effort
to develop techniques to elminate transit candidates with further
observations. Out-of-eclipse variations \citep{sir2003}, V-shaped
vs. U-shaped transits \citep{pon2005}, stellar density diagnostics
\citep{sea2003}, and the Tingley-Sackett diagnostic \citep{tin2005},
to name a few, have all contributed. Combined, these technique have
enabled CoRoT's relatively high success rate compared to, for example,
the first OGLE run, which found 2 planets among 59 equally-prioritized
candidates \citep{uda2002a,uda2002b}.  Even so, many candidates defy
easy characterization. CoRoT-7b \citep{corot7b}, to use the lightest
known transiting exoplanet as an example, required more than 100
observations with HARPS \citep{que2009}, totalling over 70 hours of
observing time.  While the host star is fairly active in this case,
thus complicating attempts to detect the radial velocity signal of the
planet, the signal of CoRoT-7b was more than an order of magnitude
larger than what one would expect for an Earth twin (3.3 m/s vs. 0.1
m/s). With the anticipated influx of candidates consistent with
terrestrial exoplanets from both CoRoT and Kepler (and potentially
future missions, such as PLATO), as well as from ground-based searches
for transits across low-mass stars, any technique that could eliminate
even a few terrestrial-class candidates would potentially save an
enormous amount of valuable telescope resources.

One avenue of candidate screening that has not been fully explored is
the density diagnostic first presented by \citet{sea2003}. Here, the
authors demonstrated that it was possible to extract stellar densities
by fitting the transit light curve with a simple trapezoid, which we
will refer to in this paper as $\rho_{\rm SMO}$. At the time that
paper was written, the required photometric measurements (5 mmag
precision with 5 minute sampling covering two transits) was not at all
typical for transit candidates -- in fact, it was quite unusual. This
paper was the first in what has turned into a long discussion on what
exactly could be done with detailed analyses of
transits. \citet{tin2005} proposed a technique to identify the best
transit candidates in a sample using only transit periods, durations,
and depths and included a discussion of the impact of unknown
eccentricity on their technique. \citet{bar2007} and \citet{bur2008}
later expanded upon this secondary analysis to conclude that transits
are more likely to occur in the elliptical orbits and this results in
higher yields from transit surverys, respectively. \citet{tres2}
revisited one part of the \citet{sea2003} derivation, stating that it
was possible to obtain stellar densities from transit fits and that
these could be used to aid in the derivation of stellar parameters
from evolutionary tracks. \citet{for2008} showed that it was possible
to estimate unknown eccentricites through an analysis of the transit
shape; \citet{kip2008} did something similar, deriving an entirely new
equation that involved no potentially erroneous
approximations. \cite{yee2008} worked to estimate period for transit
candidates with but a single transit. \citet{kip2010} compared the
accuracy of different equations for transit duration, assessing the
effect of incorrectly assuming circular orbits. \citet{bro2010}
discussed the use of stellar densities from transits to restrict
stellar radii, which are crucial in determining planetary radii.

In this paper, we combine and expand upon these ideas and propose a
slightly different technique to identify good transit candidates:
comparing densities calculated from transits to densities calculated
by other, independent means, such as $J-K$ colors, spectra in
combination with evolutionary tracks, and asteroseismology. It is our
feeling that this will be of particular interest for shallow
candidates discovered using ultra-high precision photometry from
space-based mission such as CoRoT, Kepler and potential future
missions such as PLATO for which the expected radial velocity signal
would be weak and difficult to detect.  To evaluate this technique, we
use both the set of CoRoT candidates and the known transiting
exoplanets to demonstrate that this diagnostic has the potential to be
very effective in the pre-selection of the best candidates. To aid in
this process, we derive a new equation for calculating stellar
densities using other, more tangible transit parameters often
available in papers that quote neither $a/R_\star$ nor the stellar
density.  This equation includes orbital parameters (eccentricity and
angle of periastron), which make it possible to analyze the impact of
an unknown orbital eccentricity on the measured stellar density.  It
is based on the equations in \citet{sac1999} and \citet{tin2005} and
may also be used as a consistency check of fitted transit parameters.

In $\S$~\ref{equations}, we derive the equations that will be used for
the analysis. In $\S$~\ref{independent}, we discuss the independent
density measures that compliment the stellar densities from transit
parameters. In $\S$~\ref{planets}, we apply these equations to those
transiting exoplanets for which the necessary parameters have been
published. In $\S$~\ref{corot}, we examine the use of the $\rho_{\rm
  SMO}$ combined with the densities derived from $J-K$ colors on the
CoRoT candidates. In $\S$~\ref{eccentricity}, we analyze the impact of
eccentricity on this technique. Lastly, we discuss our conclusions in
$\S$~\ref{conclusions}.

\section{Equations\label{equations}}

Calculating the stellar densities from transit parameters is not
difficult.  The most straightforward way begins with the equation for
the density of a spherical star:

\begin{equation}
\rho_\star = \frac{3 M_\star}{4 \pi R_\star^3}
\label{dens}
\end{equation}
where $\rho_\star$ is the density of the star and $M_\star$ is the
mass of the star. Then, one can use Kepler's 3rd law to substitute out
$M_\star$, needing only the rather safe assumption that $M_\star \gg
M_p$, obtaining an equation for the density of the star based on the
transit parameter $a/R_\star$ ($\rho_{t1}$):

\begin{equation}
\rho_{t1} = \frac{3\pi}{G T ^2}\left(\frac{a}{R_\star}\right)^3
\label{rho1}
\end{equation}
where $G$ is the gravitational constant and $T$ is the orbital period
of the planet \citep{sea2003,tres2}. Despite the absense of parameters
associated with orbital eccentricity, this equation does yield
accurate stellar densities from transits of planets in elliptical
orbits. Either the eccentricity and argument of the periastron are
supplied from the analysis of the radial velocity observations and
used during the transit fit or the radial velocity observations and
the photometry are fit simultaneously.

It must be pointed out that $\rho_{t1}$ is distinctly different from
$\rho_{\rm SMO}$, despite the fact that each are derived in the same
article. $\rho_{t1}$ uses parameters extracted from a full transit fit
with limb darkening to estimate the stellar density, while $\rho_{\rm
  SMO}$ uses a trapezoid fit to estimate the stellar density,
neglecting limb darkening. Moreover, $\rho_{\rm SMO}$ assumes circular
orbits, while $\rho_{t1}$ in principle takes into account eccentricity
and the argument of the periastron.

It is possible to use $a/R_\star$ as a free parameter when fitting a
transit, but many papers describing transit observations do not
include it. Moreover, one cannot intuit whether the value makes sense
from looking at a transit, which is clearly a problem in a few
cases. Therefore, it is instructive and useful to derive equations for
the stellar density and for $a/R_\star$ that depend on transit
parameters that are tangible and more commonly published. Starting
with Eq.~7 of \citet{tin2005} for the duration ($\tau_{14}$) of a
planetary transit (itself derived from equations in \citet{sac1999}):

\begin{equation}
\tau_{14}=2(R_\star+R_p)\frac{\sqrt{1-e^2}}{1+e\sin\omega}
         \sqrt{1-\frac{a^2(1-e^2)^2\cos^2 i}{(R_\star+R_p)^2(1+e\sin\omega)^2}}
         \left(\frac{T}{2\pi GM_\star}\right)^{\frac{1}{3}}
\end{equation}
where $R_p$ is the radius of the planet, $e$ is the planet's orbital
eccentricity, $\omega$ is the argument of periastron, and $i$ is the
orbital inclination. It assumes that the mass of the planet is
negligible, the star-planet separation during transit is much greater
than the stellar radius (which is not always the case), and the
orbital velocity is constant during the transit. From inspection, it
is evident that the right-hand side of above equation goes as $R_*
M_*^{-1/3}$, which is proportional to $\rho_*^{-1/3}$.  If we solve
for $M_*^{1/3}/R_*$, cube everything and multiply by $4\pi/3$, we
obtain another equation for the density ($\rho_{t2}$) of the parent
star, based on more fundamental parameters than $a/R_\star$:

\begin{equation}
\label{rho2}
\rho_{t2}= \frac{3T Q^3}{\pi^2 G \tau_{14}^3}
            \left(\frac{(1+k)^2-b^2}{1-e^2}\right)^\frac{3}{2}
\end{equation}
where the ratio of the radii $k = R_p/R_\star$, the impact parameter
$b = Qa\cos i/R_\star $ and 
\begin{equation} 
Q = \frac{1-e^2}{1+e\sin\omega},
\end{equation}
which equals 1 for circular ($e=0$) orbits. One can also solve the same
equation for $a/R_\star$, getting:

\begin{equation}
\label{eqa_r}
\frac{a}{R_\star} = \frac{T Q}{\pi\tau_{14}\sqrt{1-e^2}}
                   \sqrt{(1+k)^2-b^2}.
\end{equation}
Plugging this equation into \eqsmo yields the same results as given in
\eqmy. As this equation is invertible, transit durations can be
calculated from $a/R_\star$ as well. This makes it possible to check
the quality of a fit in cases where all of the pertinent parameters
($T$,$\tau_{14}$, $a/R_\star$, $b$, $k$, $e$, $\omega$) are
given. Additionally and significantly, this equation makes it possible
to evaluate the potential impact of an unknown orbital eccentricity on
densities determined from transit parameters (referred to generically
as transit densities or $\rho_{t}$ hereafter).

With these two equations in hand, we can assess if the assumption of
$a \gg R_\star$ is an issue -- indeed \citet{kip2010} specifically
calls into question the accuracy of this equation -- but given the
level of uncertainty in transit parameters, it ultimately make little
discernable difference. HAT-P-7b has one of the smallest $a/R_\star$s,
only 3.82 \citep{win2009a}. Even for this case, the impact on the
density is very small: $\rho_{t1} = 0.217$ and $\rho_{t2} = 0.204$ --
only 6\%, which is approximately one-third the size of the measurement
errors and not an atypical difference between $\rho_{t1}$ and
$\rho_{t2}$ for the transiting exoplanets for which both values are
available. However, planets may someday be discovered transiting very
closely to evolved or very hot stars -- for such cases, this
assumption might very well prove problematic.

\section{Independent Stellar Density Measures\label{independent}}

In order to accomplish our stated goal of evaluating transit
candidates by their stellar densities, it is necessary to have an
independent measure of stellar density to compliment the transit
density.  In the following analyses, we pursue three different
possibilities: 1) from published values of stellar radii and masses
{\it not based on transits}, 2) from $J-K$ colors, and 3) from
asteroseismology. As a group, these will hereafter be referred to as
stellar densities.

\subsection{Stellar Densities from  Published Stellar Masses and Radii}

It is a trivial matter to calculate stellar densities from the
published stellar masses and radii (referred to hereafter as
$\rho_{spec}$). These are generally derived from spectral analysis
combined with modeled evolutionary tracks (see \citet{tres2} for a
detailed discussion). Spectral analysis yields a variety of
information, such as temperature, metallacity, $\log g$ (where $g = G
M_*/R_*^2$), which can then be used with either a true distance
measure (generally from Hipparcos) or evolutionary tracks to obtain
the stellar radii and densities, using $\rho_*/\rho_\odot =
g_*/g_\odot R_*$. According to \citet{tres2}, it is possible to use
$\log g$ to constrain the stellar parameters from the evolutionary
tracks, but surface gravities from spectra generally are not very
precise, leading to highly uncertain stellar radii. They then
proposed using transit densities to improve the determination
of stellar parameters. This would suggest that at least some groups
have used the transit densities to help determine their stellar
parameters; while it was not done before the TrES-2 paper
\citep{tres2}, it has in the intervening years become essentially
universal. Therefore, we caution that for some published transiting
exoplanets, this density ($\rho_{spec}$) may not be truly independent of
the transit density.

\subsection{Stellar Densities from $J-K$  colors}

\begin{figure}
\plotone{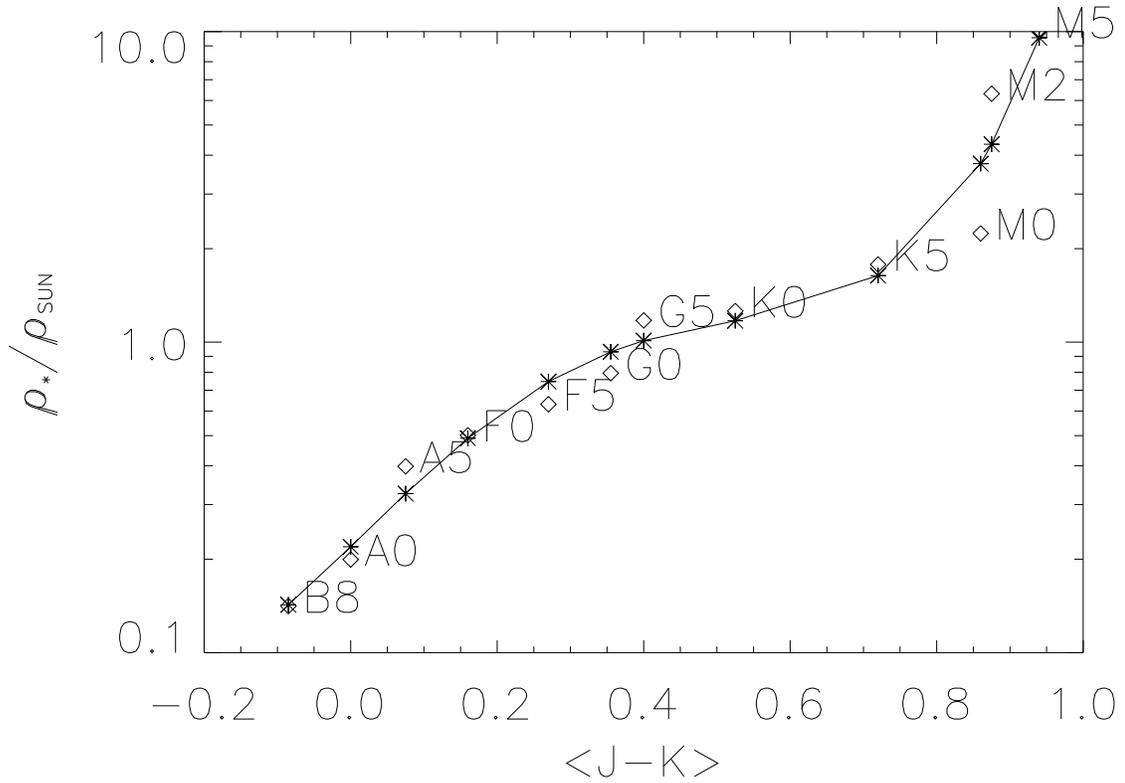}
\caption{$J-K$ colors and spectral types vs. stellar density. This
  figure shows how the stellar density varies with $J-K$ color and
  spectral type for the main sequence. Diamonds show the actual values
  from Allen's Astrophysical Quantities \citep{cox00}, while the line
  and asterisks show the interpolated values (a fourth-degree
  polynomial fit) used to calculate the $J-K$ stellar densities used
  in this paper.
\label{jkdens}}
\end{figure}

Allen's Astrophysical Quantities \citep{cox00} lists $J-K$ colors and
stellar masses and radii as a function of spectral type for main
sequence stars. While these values are very general, being independent
of factors such as age and metallicity, they may offer a reasonable
approximation of stellar density, at least for the purpose of
identifying less-than-ideal candidates. The advantage of this measure
is that $J-K$ colors are readily available for most stars from the
2MASS database \citep{skr06}, which means that no additional
observations or analysis are necessary -- simply look up the $J-K$
color, then interpolate among the values from Allen's Astrophysical
Quantities to arrive at the density ($\rho_{JK}$) considering only
main sequence stars. This is illustrated in Fig.~\ref{jkdens}, which
plots the stellar density versus the $J-K$ color.

Apparently, this measure has several weaknesses. First, it assumes
that all stars are main sequence stars. Obviously, this is not always
the case; stars evolve, and once off the main sequence the resulting
$J_K$ densities would be incorrect.  Second, while $J-K$ is a very red
color, stars residing in dust-rich environs can experience enough
interstellar reddening to have a significant effect on their resulting
density. For example, CoRoT-10b has color excess $E(J-K) = 0.24$
\citep{corot10b}, which leads to a difference in
$\log(\rho_\star/\rho_\odot)$ of about 0.5 dex -- or a factor 1.6.

Technically speaking, the 2MASS $J$ and $K$ colors are slightly
different than the Bessel and Brett \citet{bes1988} system used in
Allen's Astrophysical Quantities. Therefore, to be completely
rigorous, a conversion from the one system to the other ought to be
performed.  However, such a conversion results in a very small
change in stellar density -- only a few hundreds of a solar
density. Considering the magnitude of the generalizations involved in
this density measure, this difference can be neglected.

\subsection{Asteroseismic Densities}

Another technique exists that can reliably obtain stellar densities,
though we cannot exploit it yet in this work: asteroseismology. By
measuring the frequencies of the acoustic modes and performing some
model-fitting, it is possible to get a good measure of the stellar
density from the average frequency spacing between these modes in the
asymptotic region \citep{van1967, tas1980}. It is only in the last few
years that these measurements have become possible, generally through
the radial velocity fingerprint of these oscillations
(\citet{kje2008}). Recently, however, circumstances have changed: the
Kepler satellite mission has the capacity to obtain photometric time
series with the necessary precision to reveal stellar oscillations,
as would proposed missions in the future such as PLATO.
This is potentially very interesting: the photometric precision
necessary to discover transiting terrestrial exoplanets should also be
sufficient to detect oscillations, thereby offering a sensitive
density measure without additional observation.

\section{Application to Known Transiting Exoplanets\label{planets}}
We performed an extensive literature search to gather the transit
parameters for as many transiting exoplanets as possible.
Table~1 lists the planets and their periods, along
with their corresponding $\rho_{t1}$ and $\rho_{t2}$ (calculated from
the available transit parameters) and $\rho_{JK}$, when possible --
some faint (or very crowded) candidates do not appear in the 2MASS
database. Asteroseismic densities are much less common, however: only
a single recent paper by \citet{chr2010} publishes asteroseismic
densities ($\rho_a$), and then only for the three stars known to
harbor exoplanets in the Kepler field before the mission began:
HAT-P-7, HAT-P-11, and TrES-2 -- the last two with preliminary stellar
parameters only. These are also included in
Table~1.

\begin{figure}
\plottwo{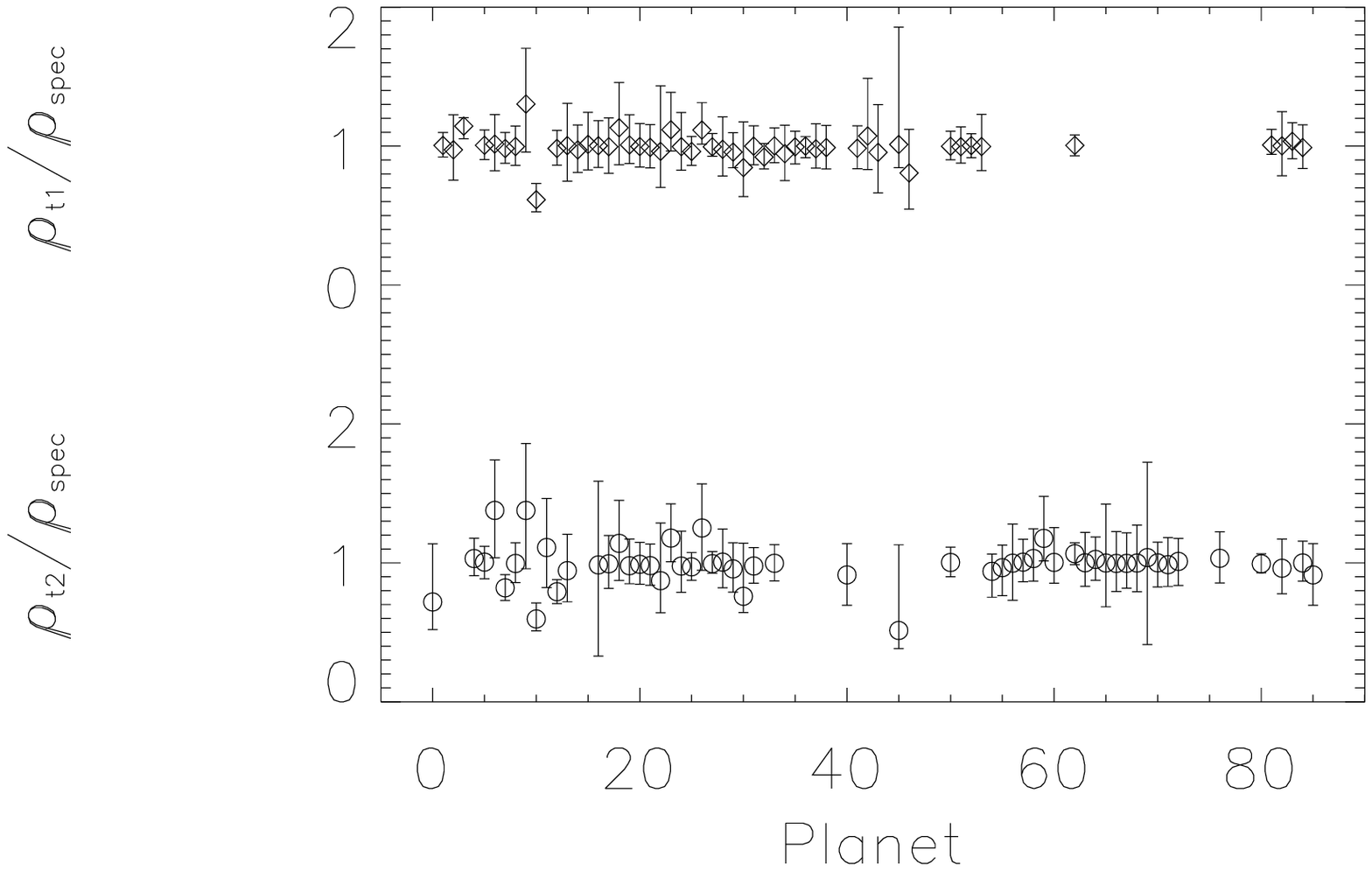}{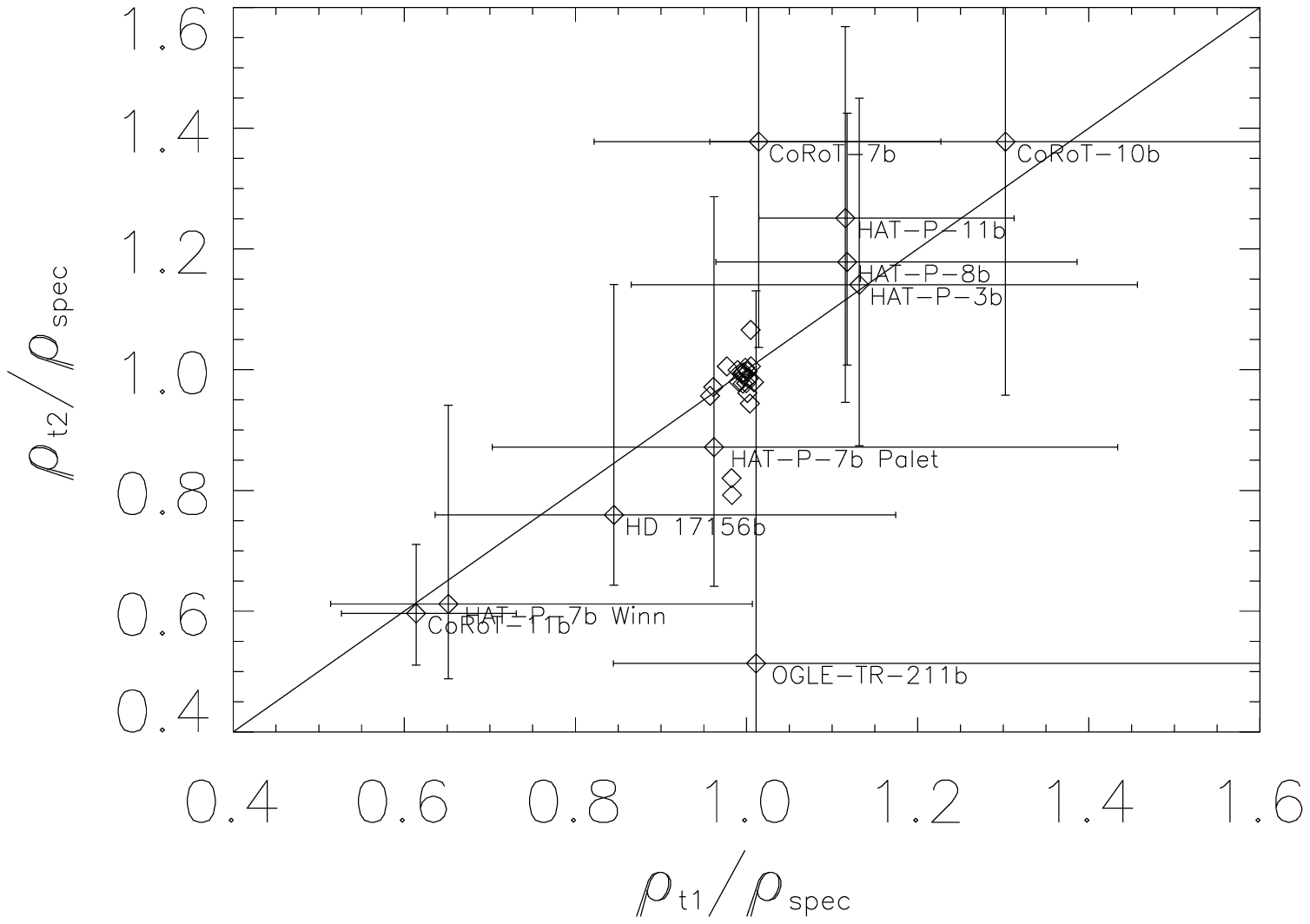}
\caption{The left-hand figure shows the density ratios for the known
  transiting exoplanets for which the necessary information is
  available. Density ratios are created using spectra ($\rho_\star$)
  and either $\rho_{t1}$ (top) or $\rho_{t2}$ (bottom). Notice how
  consistently these values are centered around 1, particularly for
  $\rho_{t1}$ -- clear evidence of the widespread use these
  densities have in the final determination of stellar parameters. The
  right-hand figure reiterates this with a comparison of the density
  ratios calculated using $\rho_{t1}$ and $\rho_{t2}$ for the subset of
  transiting exoplanets for which all the necessary parameters are
  published. Note that the values for $\rho_{t1}$ and $\rho_{t2}$ are
  very similar in most cases, even for planets with high
  eccentricities, verifying that $\rho_{t2}$ is accurate.  Those that
  appear to be outliers actually aren't -- the errors bars are large
  enough to include the anticipated result. Those that deviate from
  ratios of 1 but remain close to the overplotted line probably have
  problems obtaining the stellar parameters from the spectra (and
  models), while those that are displaced upwards and downwards
  typically have problems with the photometry.
\label{fig_plandens}}
\end{figure}

Fig.~\ref{fig_plandens} shows the ratio of the transit densities and
the stellar densities from masses and radii, with $\rho_{t1}$ on the
top and $\rho_{t2}$ on the bottom. These figures demonstrate that, for
nearly all of the candidates, the two values agree within the errors
bars -- even the eccentric ones. If anything, they agree far more than
we would expect from the size of the error bars -- evidence that the
practice of using transit densities to constrain model fits of stellar
parameters is widespread. Only four planets seem to have deviations
greater than $1\sigma$ from a density ratio of one for either
$\rho_{t1}/\rho_{spec}$ or $\rho_{t2}/\rho_{spec}$: HAT-P-7b,
CoRoT-11b, CoRoT-7b and OGLE-TR-211b. The first case is resolved when
the transit parameters are taken from the same reference as the
stellar parameters \citep{pal2008}, instead of a later paper
\citep{win2009a}, which contains better photometry and consequently
more precise transit parameters but does not repeat the spectral
analysis of the parent star. This presents a clear case where the
derived transit density influences the stellar parameters.  The second
offers no easy explanation and requires further study.  The last two
do not have sufficiently precise photometry for confident parameter
estimation: CoRoT-7b has an extremely shallow transit and it has been
further hypothesized that magnetic activity or transit timing
variations may affect the shape of the transit \citep{corot7b} and
OGLE-TR-211b has not been observed well enough -- several different
transits are spliced together to create the final light curve,
introducing undefined uncertainties. This emphasizes the fact that
good photometry without significant unknown systematics is essential
for the use of this technique.

\begin{figure}
\plotone{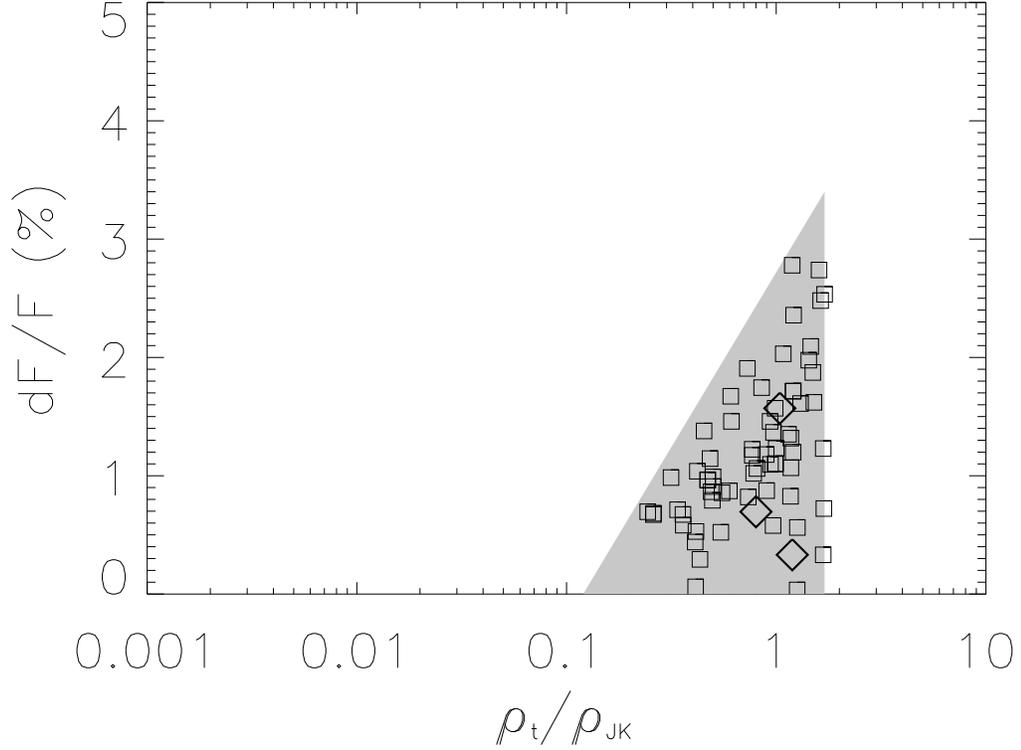}
\caption{$\rho_{transit}/\rho_{\star}$ vs. transit depth for known
  exoplanets. This figure shows the density ratios (in this case,
  $\rho_{t1}/\rho_{JK}$, when available, otherwise
  $\rho_{t2}/\rho_{JK}$) for all the known transiting exoplanets
  (squares) and for the three cases for which the asteroseismic
  density ($\rho_{t1}/\rho_a$ or $\rho_{t2}/\rho_a$ is known
  (diamonds)). The depth is approximated as the square of the
  planet-star radius ratio. The known exoplanets have density ratios
  distributed in a way that appears to be asymmetric relative to the
  ideal $\rho_{transit}/\rho_{\star} = 1$ value, likely due to
  stellar evolution (shaded region). The limited sample of
  asteroseismic densities appears to be more reliable, agreeing
  well with the transit densities regardless of transit depth.
  \label{densdep3}}
\end{figure}

As can be seen in Fig.~\ref{densdep3}, the details of the distribution
of the density ratios is interesting.  The region is largely
triangular, with a narrower distribution for deep transits and wider
for shallower ones. This apparent dependency could arise from two
causes: underestimations of $\rho_{t1}$ and $\rho_{t2}$ or
overestimation of $\rho_{JK}$. It is unlikely that bad transit
parameters are the source of this problem: while shallower transits
would tend to have less precise transit parameters, the precision of
the transit parameters for the CoRoT planets is more limited by
uncertainties in the limb darkening parameters than photometric
precision and they exhibit the same tendency. Therefore, we suspect
that $\rho_{JK}$ is the source. Most of the planets -- but of course
not all -- discovered are approximately Jupiter-sized.  Therefore,
some fraction of shallower transits correspond to larger (evolved)
stars. The $\rho_{JK}$s derived are based on the assumption that the
stars are still on the main sequence. If the $J-K$ colors accurately
represent the densities of main sequence stars, some of the larger
stars will be more evolved stars, rather than higher mass, explaining
the presence of unexpectedly low density ratios, leading to the
observed increase in the spread in density ratios as transit depth
decreases. The numbers actually bear this out; the four exoplanets
with the lowest density ratios (HAT-P-4b, HAT-P-13b, HD 149026b, and
HD 17156b) have orbit stars with effective temperatures between 5650
and 6150 Kelvin, but $R\sim 1.5R_\odot$ and $\log g \sim 4.2$, when
$\sim 4.5$ is more typical for main sequence stars with these
temperatures -- indeed, HD 149026b is classified as a subgiant.

Asteroseismic densities may be more accurate than $\rho_{JK}$, the few
known examples exhibiting little scatter around the expected density
ratio of unity. While the current sample is too small for conclusions,
we note that additional observations are not necessary to obtain the
asteroseismic density, only some analysis. Therefore, a comparison
between asteroseismic densities and transit densities, when possible,
should prove to be an extremely valuable diagnostic tool in evaluating
transit candidates. This may be particularly useful for
candidates for terrestrial exoplanets in multi-planet systems -- some
groups have hypothesized that terrestrial planets may favor
low-eccentricity orbits to maintain stability, both with giant planets
present \citep{pil2009} and without \citep{man2010}. This topic is still
being debated, however, and does not address single planet systems al all.

\begin{figure}
\plotone{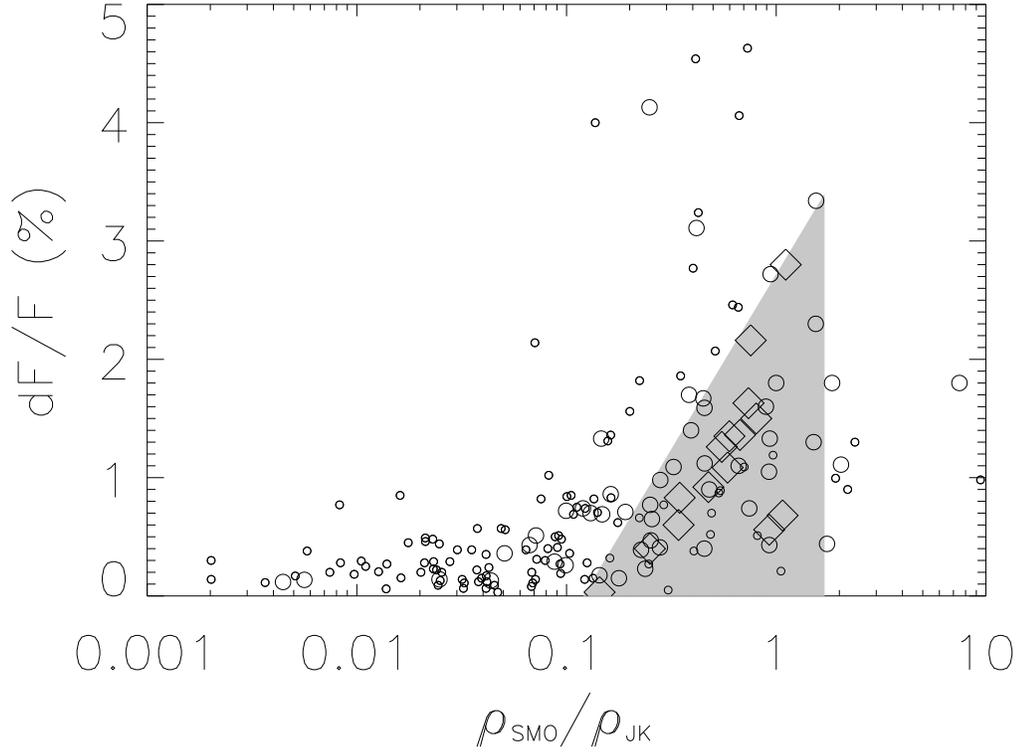}
\caption{$\rho_{transit}/\rho_{\star}$ vs. transit depth plot for the
  CoRoT candidate list. Diamonds are the confirmed exoplanets, while
  large circles are the highest priority CoRoT candidates and small circles
  lower-priority CoRoT candidates. For these, $\rho_{\rm SMO}$ is the
  transit density estimated by the trapezoid method and $\rho_{JK}$ is
  the stellar density. The shaded region is the same as in
  Fig.~\ref{densdep3}. Notice how the CoRoT planets fall in the same
  shaded region. Also notice that the vast majority of candidates have
  very low density ratios, while very few have ratios higher than
  1. This is because candidates in both giant stars and significantly
  blended stars will have low density ratios \citep{sea2003}, while
  only eccentricity or errors can increase density ratios. This
  demonstrates how effectively density ratios can be for screening
  transit candidates. It also shows that $\rho_{\rm SMO}$ is a
  reasonable proxy for $\rho_{t1}$ and $\rho_{t2}$.
  \label{densdep1}}
\end{figure}


\begin{figure}
\plotone{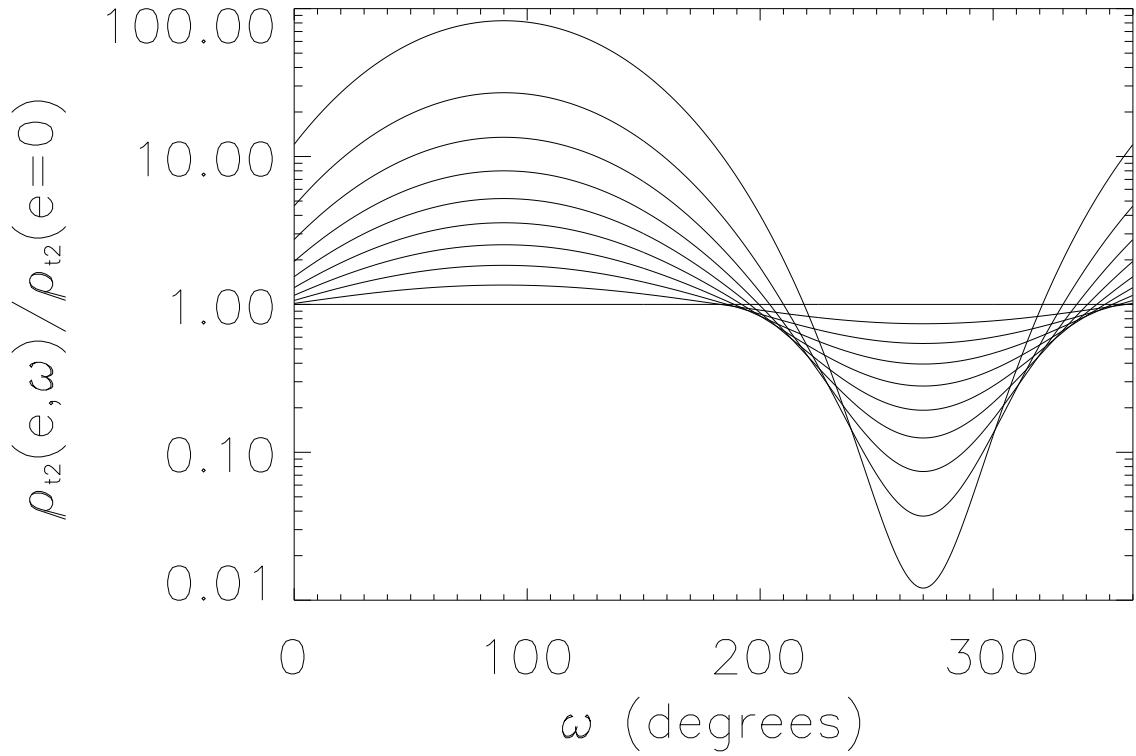}
\caption{Density Ratio vs. eccentricity and angle of periastron. This
  figure shows how the density ratio varies as a function of angle of
  periastron and eccentricity. The lines from outermost to innermost
  depict eccentricities going from 0.9 to 0 in steps of 0.1. Notice
  how the angle of periastron produces density ratios greater than 1
  over a larger range of values than it does for ratios less than
  1. \label{fig_densew}}
\end{figure}

\begin{figure}
\plotone{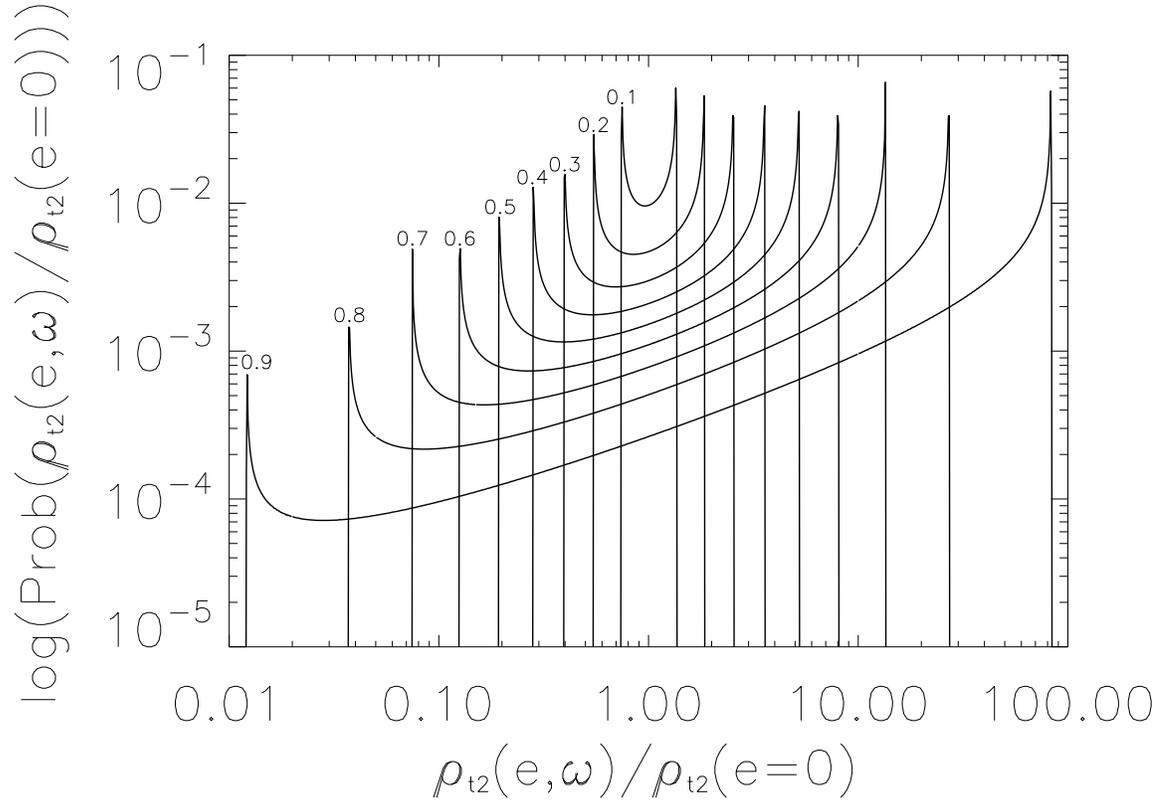}
\caption{Density ratio vs. probability and eccentricity. This figure
  shows the probability of measuring a particular density ratio as a
  function of eccentricity. The lines from innermost to outermost
  depict eccentricities going from 0.9 to 0.1 in steps of 0.1. Notice
  how the probability that the density ratio is greater than 1
  increases with eccentricity. \label{fig_densrat}}
\end{figure}

\begin{figure}
\plotone{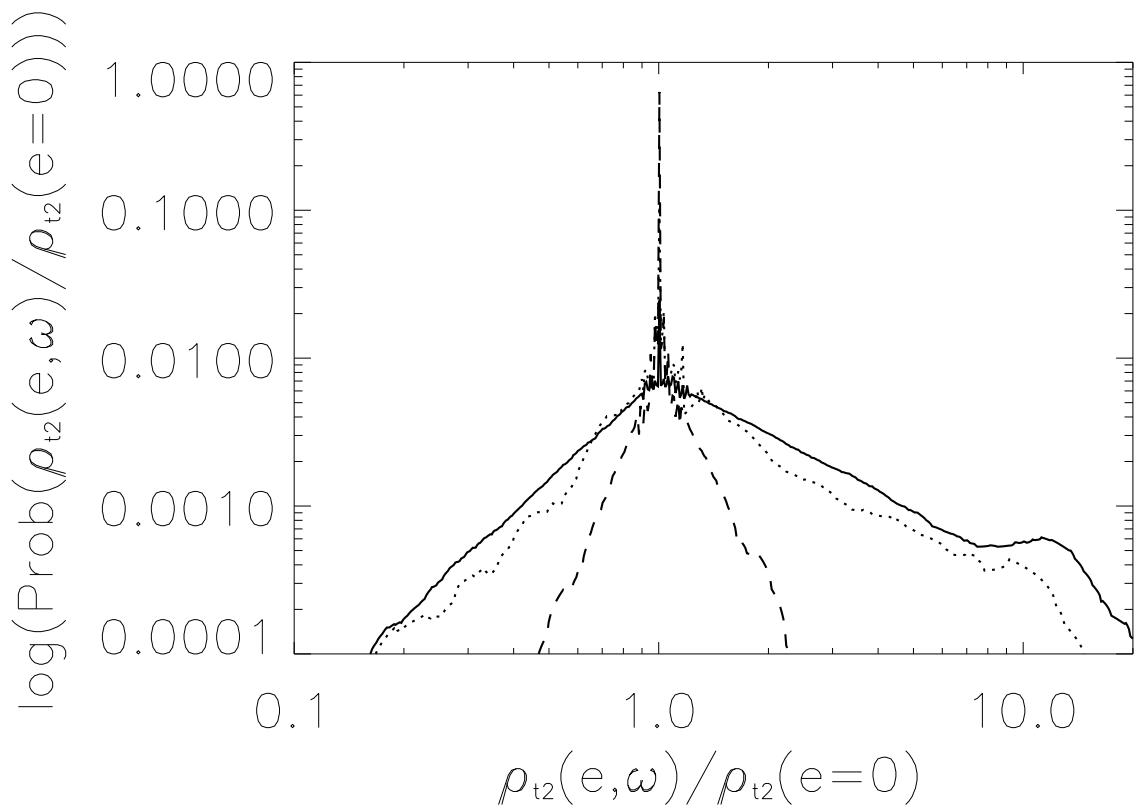}
\caption{Density ratio probabilities vs. period. This figure shows the
  probability that a transiting planet with unknown $e$ and $\omega$
  will produce a given density ratio for three different period
  regimes, using the observed distribution of exoplanet eccentricities.
  The dashed line is for planets with periods less than
  5 days, while the dotted line is for planets with periods of 5 to
  100 days and the solid line is for planets with periods longer than
  100 days. Notice how the peak at a density ratio equal to one
  decreases as period increases, while the chance for extreme density
  ratios increases. Please note that the lines have been smoothed for
  clarity.
  \label{densprob}}
\end{figure}

\begin{figure}
\plotone{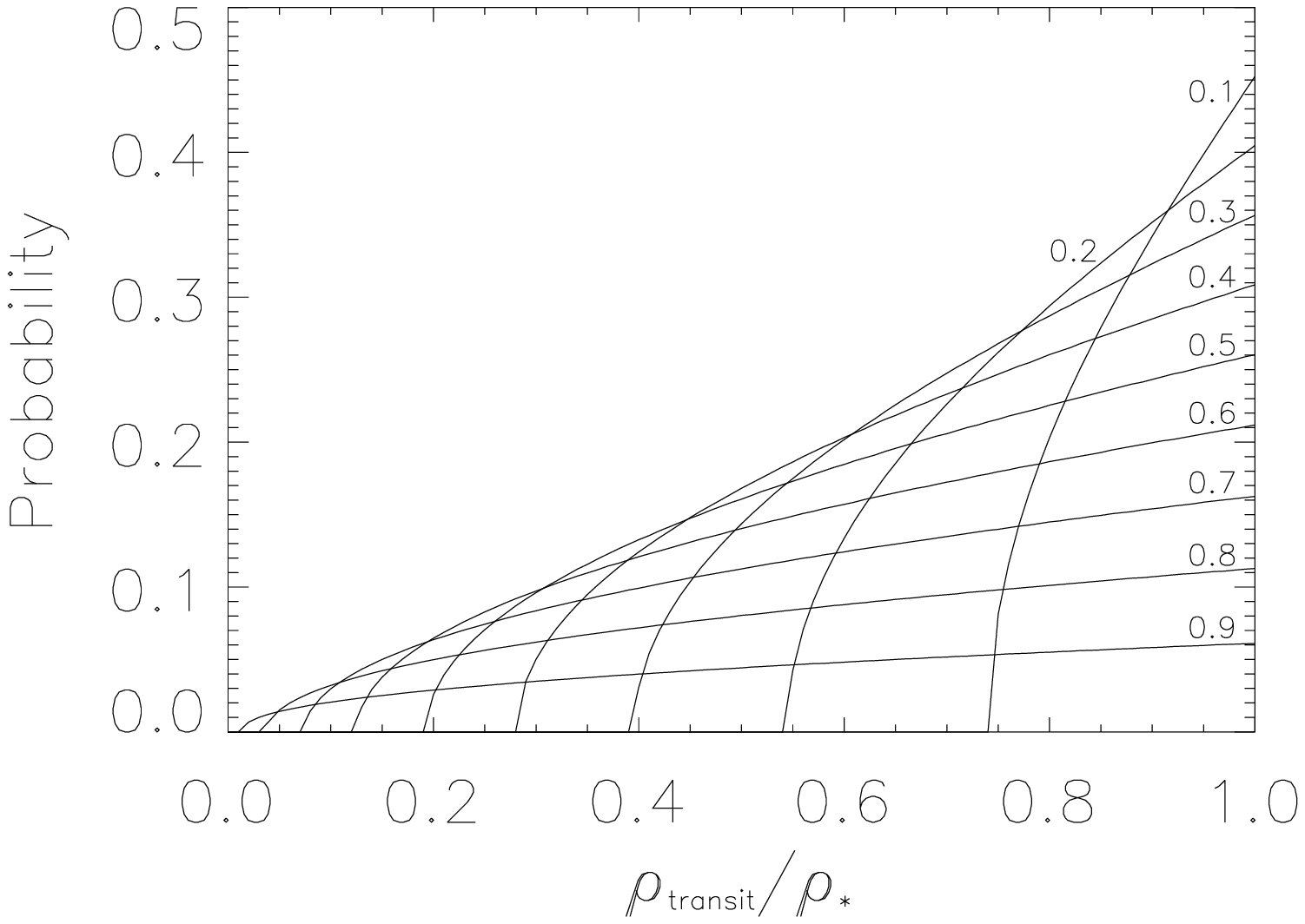}
\caption{Density ratio probabilities vs. eccentricity. This figure
  shows the probability that a given eccentricity with an unknown
  periastron angle will have a transit density {\em lower} than the
  listed value. The eccentricity ranges from 0.1 to 0.9 in steps of
  0.1. From this plot, it is clear that high eccentricities strongly
  favor transit densities greater than 1, while low-to-moderate
  eccentricities have a significant probability of producing
  comparatively low density ratios. Arbitrarily choosing a density
  ratio cut-off of $\geq 0.5$ would falsely de-prioritize $\sim$ 11\%
  of transits with an eccentricity of 0.7, , $\sim$16\% of transits
  with an eccentricity of 0.4 and {\em no} transits with an
  eccentricity less than about 0.22, while including relatively few
  astrophysical false positives. It should also be mentioned that low
  eccentricities $<0.3$ are more common, particularly for exoplanets
  with periods less than 5 days -- which transit searches are most
  likely to find.
  \label{ltprob_e}}
\end{figure}

\section{Application to CoRoT Candidates\label{corot}}

The set of CoRoT candidates offers an excellent opportunity to test
this technique. Many of these candidates are listed in papers
describing the results of individual CoRoT runs -- e.g. IRa02
\citep{cap2009} and LRc01 \citep{cab2009} -- but others are not yet
published. Ideally, we would want to use
transit densities calculated from detailed transit fits and compare
them to densities from spectra, but this information is not currently
available for more than a handfull of candidates. However, as part of
the array of information used for ranking candidates, the CoRoT
Science Team determines both $\rho_{\rm SMO}$ (derived from a
trapezoid fitting that neglects limb darkening) and $\rho_{JK}$. Each
of these techniques is somewhat less than ideal, but without spectral
information for each candidate, it is impossible to use detailed
transit fitting including limb darkening and stellar evolutionary
tracks to get better densities. They are apparently sufficient,
however, to test the potential of density for candidate evaluation. As
can be seen in Fig.~\ref{densdep1}, these simple density measures work
remarkably, perhaps even surprisingly, well. While not all of the
planets line up along the line describing the density ratio $\rho_{\rm
  SMO}/\rho_\star = 1$, most of the planets do however fall in or
are very close to the same well-defined region as the planets shown in
Fig.~\ref{densdep3}.  We note that the follow-up process is incomplete
and that some of the candidates in that region may in fact be
unconfirmed planets.

It is also clear in this figure that the majority of candidates have
very low density ratios. This can be primarily attributed to two
causes beyond the limitations of $\rho_{\rm SMO}$: giant stars and
blends, which are eclipsing binaries whose light is accompanied by a
bright third star, causing deep eclipses to be diluted and thus appear
like planetary transit.  As already mentioned in \citet{sea2003}, both
blends and giant stars will have unusually low values of $\rho_{\rm
  SMO}$ -- the giants stars because their density is truly much lower
than main sequence stars and the blends because the inclusion of the
light of the third star leads the trapezoid fit to converge on a
solution that is larger than the bright third star and with a higher
impact parameter. This overestimation of the radius leads in turn to
an underestimation of the stellar density of the third star of up to
50\%.

The known planets in this figure support the hypothesis that the
distribution of densities ratios is due more to the insufficiency of
$\rho_{JK}$ than the transit densities. The precision of CoRoT data is
extremely high, so the errors in the transit parameters are dominated
by uncertainties in the limb darkening rather than limitations of the
photometry. Indeed, despite the fact that $\rho_{\rm SMO}$ is built
upon more risky assumptions than $\rho_{t1}$ and $\rho_{t2}$ and
therefore presumably less reliable, the distribution of the $\rho_{\rm
  SMO}$ densities for the CoRoT planets are essentially
indistinguiable from the distribution of the $\rho_{t1}$ and
$\rho_{t2}$ densities for all exoplanets (including the CoRoT planets)
-- while the individual values are different, the distribution is not.
The only exception is CoRoT-7b, for which the extracted transit
parameters are likely distorted by magnetic activity or transit timing
variations. This demonstrates that $\rho_{\rm SMO}$ is actually quite
useful in and of itself (particularly for more central transits) given
the errors inherent in the extracted transit parameters, which may or
may not include impossible-to-define uncertainties in limb darkening
coefficients.

\section{Impact of Eccentricity on Densities from Transits\label{eccentricity}}

As the transit density is a function of the orbital eccentricity and
angle of periastron, the danger in using this diagnostic blindly is
the high probability of a significantly non-zero eccentricity -- most
planets have significant eccentricity, particularly among planets with
periods greater than 5 days or so. To evaluate the value of this
technique properly, it is necessary to evaluate the influence of
non-zero eccentricities. 

It is a relatively straightforward matter to use the equation for
$\rho_{t2}$ (which directly includes eccentricity while the equation
for $\rho_{t1}$ does not) to explore this dependency. By comparing the
response of $\rho_{t2}$ as $e$ and $\omega$ are changed to $\rho_{t2}$
with $e=0$, we can observe the effect of unknown eccentricity on the
$\rho_{t2}(e,\omega)/\rho_{t2}(e=0)$ density ratios. This is shown in
Fig.~\ref{fig_densew}. These can readily be converted into the
probability of observing a particular density ratio as a function of
$e$, if $\omega$ is unknown. This is shown in
Fig.~\ref{fig_densrat}. Here, we see that, as eccentricity increases,
the probability of observing a density ratio greater than 1
increases. This is the case because the probability that a transit
occurs goes as the inverse of the star-planet separation, which
therefore more likely that a transiting planet will transit when it is
close to its parent star, rather than far away. The closer the planet
is to its parent star, the higher its orbital velocity relative to its
hypothetical $e=0$ velocity. Higher velocities lead to shorter transit
durations and thus higher transit densities. The higher the
eccentricity, the more pronounced this effect: 93.9\% of all orbits
with $e=0.9$ will produce a density ratio greater than 1.  This can
then be combined with the observed distribution of planet
eccentricities as a function of period (Fig.~\ref{densprob}), to give
an idea of what fraction of planets with a given density ratio and
period are planets. It is important to remember that this sample is
comprised almost entirely of giant planets, however -- the
distribution of eccentricity as a function of period might be
different for terrestrial planets.

One final consideration must be remarked upon: while density ratios
larger than 1 are more common for planets of unknown eccentricities,
very low density ratios will also occur, albeit rarely. These will
exhibit very distinctive light curves that display very long,
flat-bottomed transits. A density ration of 0.01 would result in an
increase in the duration by a factor of $\sqrt[3]{1/0.1} = 4.6$ over
that expected for a circular orbit with the same period. Given that
planets in circular orbits with periods of a year would have transit
durations over 12 hours, such planets could have transit durations up
to several days, but would stand out among false positives with
similar density ratios. To the best of our knowledge, only eclipses of
giant stars could produce similar events and these can be quickly
identified with spectra due to their low surface gravity. For the
interpretation of transit candidates, this means that cases with low
$\rho_{t}/\rho_{JK}$ (or $\rho_{t}/\rho_{a}$, if available) should
first be studied to see if the prospective parent star is a giant.  If
not, the probability that the candidate is a planet remains very low
(see Fig.~\ref{ltprob_e}) and should thus be assigned a very low
priority.

To apply this method, we recommend selecting a lower limit for the
density ratio as a function of the period: only planets with periods
$\gtrsim 100d$ are currently known to have a chance to have very
extreme eccentricities; in fact, the maximim possible eccentricity
dereases with period: $e \lesssim 0.35 \log (T/1d) + 0.24$ (based on
figure 4 from \citet{corot9b}). Accordingly, no known 'hot' planets ($T
< 5d$) have an eccentricity higher than 0.265. With this in mind, we
can therefore conclude, per Fig.~\ref{ltprob_e}, that all candidates
with density ratios lower than 0.45 can be strongly
de-prioritized. Similar 'secure' cut-offs could be established for
candidates with longer periods, which could potentially possess larger
eccentricities. Additionally, 'semi-secure' cut-offs could be
implemented, which would have small residual probabilities of falsely
de-prioritizing very eccentric planets. However, should $\rho_{JK}$ be
used to create the density ratio, it is perhaps better to be more
circumspect given this density measure's difficulties with evolved
stars.


\section{Conclusions\label{conclusions}}

Ratios of densities derived from transit parameters over those from
$J-K$ colors (or some other independent measure, such as
asteroseismology) are a useful tool for the identification of transit
exoplanets out of a much larger set of candidates. Even rudimentary
density measures (trapezoidal and $J-K$ densities) appear to identify
exoplanets effectively among the CoRoT candidates. More precise
density measures from detailed transit fitting and stellar spectra
will be even more effective; these, however, require additional
spectral observations. Given data of sufficient quality, densities
from asteroseismology might be even more effective, although at this
time the sample of systems with transiting exoplanets and
asteroseismic densities is too small to draw a definitive
conclusion. Significant eccentricities will impact the measured
density ratio in the absense of any knowledge of these orbital
parameters; however, they will tend to increase the density ratios of
discovered transiting exoplanets, while most false positives have
density ratios lower than typical planets.  It is important to
remember that any tool can be misused; the one describe in this paper
is no exception, but it does represent an improvement on other
techniques currently used to prioritize transit candidates using
transit parameters. Even a 'secure' density cut-offs as described in
the previous section may de-prioritize some true exoplanets due to any
number of causes: extreme planet or star characteristics or otherwise
inaccurately measured density ratios, for example.  However, when used
in conjunction with other techniques (ellipsoidal variations, presence
of secondaries, etc.), density ratios can help identify the candidates
which are most likely to be exoplanets and which are most likely to be
false positives, increasing the efficiency of efforts to confirm or
reject planet candidates.


\tiny
\setlongtables
\begin{longtable}{c|c|c|c|c|c|c|c}
\multicolumn{7}{l} {\hspace{-0.9cm} \normalsize Table 1.--- Stellar densities (in ${\rm g\,cm}^{-3}$)}\\
\multicolumn{7}{l}{} \\
\hline
Planet     & reference               & $\rho_{t1}$ & $\rho_{t2}$        & $\rho_{JK}$ & $\rho_{s}$ & $\rho_a$\\
\hline
\endfirsthead
\multicolumn{7}{l}{Continued from previous page}\\
\hline
Planet     & reference               & $\rho_{t1}$ & $\rho_{t2}$        & $\rho_{JK}$ & $\rho_{s}$ & $\rho_a$\\
\hline
\endhead
\hline
\multicolumn{7}{l}{Continued on next page}\\
\endfoot
\hline
\endlastfoot
\hline                               
CoRoT-1b   & \citet{corot1b}   &       $0.87^{+0.40}_{-0.09}$ &                      & 1.19 & $1.21^{+0.42}_{-0.31}$ & \\ 
CoRoT-2b   & \citet{corot2b}   & $1.87^{+0.03}_{-0.03}$       &                      & 1.57 & $1.87^{+0.16}_{-0.16}$ & \\ 
CoRoT-3b   & \citet{corot3b}   & $0.50^{+0.08}_{-0.07}$       &                      & 1.21 & $0.51^{+0.10}_{-0.09}$ & \\
CoRoT-4b   & \citet{corot4b}   & $1.17^{+0.05}_{-0.01}$       &                      & 1.25 & $1.02^{+0.09}_{-0.03}$ & \\
CoRoT-5b   & \citet{corot5b}   &       $0.87^{+0.08}_{-0.06}$ &                      & 1.43 & $0.85^{+0.09}_{-0.08}$ & \\
CoRoT-6b   & \citet{corot6b}   &       $1.38^{+0.08}_{-0.12}$ & $1.38^{+0.08}_{-0.07}$ & 1.43 & $1.38^{+0.13}_{-0.12}$ & \\ 
CoRoT-7b   & \citet{corot7b}   &       $2.75^{+0.60}_{-0.57}$ & $2.02^{+0.30}_{-0.27}$ & 1.61 & $1.99^{+0.31}_{-0.26}$ & \\ 
CoRoT-8b   & \citet{corot8b}   &       $2.67^{+0.19}_{-0.18}$ & $2,23^{+0.15}_{-0.16}$ & 2.12 & $2.72^{+0.26}_{-0.24}$ & \\
CoRoT-9b   & \citet{corot9b}   &       $1.68^{+0.17}_{-0.16}$ & $1.68^{+0.17}_{-0.16}$ & 1.43 & $1.68^{+0.18}_{-0.17}$ & \\
CoRoT-10b   & \citet{corot10b} &       $3.32^{+0.73}_{-0.64}$ & $3.51^{+0.97}_{-0.86}$ & 2.54 & $2.55^{+0.57}_{-0.45}$ & \\
CoRoT-11b   & \citet{corot11b} &       $0.69^{+0.02}_{-0.02}$ & $0.67^{+0.03}_{-0.03}$ & 1.43 & $1.13^{+0.18}_{-0.18}$ & \\
CoRoT-12b   & \citet{corot12b} &                              & $1.22^{+0.13}_{-0.14}$ & 1.43 & $1.10^{+0.33}_{-0.25}$ & \\
CoRoT-13b   & \citet{corot13b} &       $1.47^{+0.13}_{-0.13}$ & $1.18^{+0.07}_{-0.07}$ & 1.26 & $1.49^{+0.14}_{-0.13}$ & \\
CoRoT-14b   & \citet{corot14b} &       $0.90^{+0.17}_{-0.15}$ & $0.85^{+0.12}_{-0.12}$ & 1.64 & $0.09^{+0.22}_{-0.17}$ & \\
GJ 436b    & \citet{gj436b}    &                            & $6.81^{+0.78}_{-0.72}$ & 4.05 & $7.00^{+1.04}_{-0.88}$ & \\
GJ 1214b   & \citet{gj1214b}   &                            & $23.9^{+2.1}_{-1.9}$  & 20.9 & $23.6^{+4.6}_{-4.1}$ & \\ 
HAT-P-1b   & \citet{hat1b}     &       $1.12^{+0.67}_{-0.73}$    & $1.14^{+0.11}_{-0.10}$ & 1.15 & $1.14^{+0.17}_{-0.15}$ & \\
HAT-P-2b   & \citet{hat2b}     &       $0.43^{+0.06}_{-0.05}$    & $0.43^{+0.06}_{-0.06}$ & 0.80 & $0.44^{+0.07}_{-0.06}$ & \\
HAT-P-3b   & \citet{hat3b}     &       $2.69^{+0.53}_{-0.53}$    & $2.67^{+0.58}_{-0.53}$ & 1.60 & $2.36^{+0.34}_{-0.37}$ & \\
HAT-P-4b   & \citet{hat4b}     &       $0.434^{+0.002}_{-0.002}$ &  $0.446^{+0.039}_{-0.007}$ & 1.25 & $0.44^{+0.07}_{-0.07}$ & \\
HAT-P-5b   & \citet{hat5b}     &       $1.02^{+0.06}_{-0.07}$ & $1.03^{+0.08}_{-0.08}$ & 1.34 & $1.03^{+0.15}_{-0.13}$ & \\ 
HAT-P-6b   & \citet{hat6b}     &       $0.57^{+0.04}_{-0.04}$ & $0.58^{+0.05}_{-0.05}$ & 0.97 & $0.59^{+0.08}_{-0.07}$ & \\
HAT-P-7b   & \citet{pal2008}   &       $0.29^{+0.06}_{-0.06}$ & $0.32^{+0.08}_{-0.07}$ & 0.89 & $0.33^{+0.07}_{-0.10}$ & $0.2712 \pm 0.0032$\\ 
           & \citet{win2009a}  &       $0.20^{+0.07}_{-0.02}$ & $0.22^{+0.07}_{-0.03}$   & 0.89 & $0.33^{+0.07}_{-0.10}$ & \\
HAT-P-8b   & \citet{hat8b}     &       $0.54^{+0.07}_{-0.05}$ & $0.51^{+0.09}_{-0.04}$ & 1.03 & $0.46^{+0.06}_{-0.07}$ & \\
HAT-P-9b   & \citet{hat9b}     &       $0.77^{+0.08}_{-0.07}$ & $0.78^{+0.06}_{-0.01}$ & 1.02 & $0.79^{+0.16}_{-0.15}$ & \\
HAT-P-10b  & \citet{hat10b}    &       $2.31^{+0.14}_{-0.13}$ & $2.28^{+0.16}_{-0.15}$ &      & $2.38^{+0.21}_{-0.19}$ & \\ 
HAT-P-11b  & \citet{hat11b}    &       $3.36^{+0.79}_{-0.77}$ & $3.00^{+0.45}_{-0.10}$ & 1.79 & $2.69^{+0.25}_{-0.23}$ & $2.5127 \pm 0.0009$\\
HAT-P-12b  & \citet{hat12b}    &       $2.99^{+0.11}_{-0.13}$ & $2.99^{+0.16}_{-0.12}$ & 2.10 & $3.00^{+0.18}_{-0.22}$ & \\
HAT-P-13b  & \citet{hat13b}    &       $0.46^{+0.06}_{-0.05}$ & $0.44^{+0.06}_{-0.06}$ & 1.31 & $0.45^{+0.08}_{-0.07}$ & \\ 
HD 149026b & \citet{149026b}   &       $0.50^{+0.09}_{-0.07}$ & $0.50^{+0.05}_{-0.04}$ & 1.15 & $0.52^{+0.05}_{-0.04}$ & \\ 
HD 17156b  & \citet{17156b}    &       $0.45^{+0.20}_{-0.03}$ & $0.50^{+0.16}_{-0.10}$ & 1.21 & $0.59^{+0.09}_{-0.11}$ & \\ 
HD 189733b & \citet{189773b}   &       $2.65^{+0.20}_{-0.200}$ & $2.71^{+0.25}_{-0.24}$ & 1.67 & $2.71^{+0.31}_{-0.27}$ & \\ 
HD 209458b & \citet{southw}    &                           & $1.025^{+0.006}_{-0.006}$ & 1.10 & $1.11^{+0.11}_{-0.11}$ & \\ 
HD 80606b  & \citet{80606b}    &       $1.63^{+0.15}_{-0.15}$ & $1.64^{+0.14}_{-0.14}$ & 1.40 & $1.63^{+0.16}_{-0.14}$ & \\
Kepler-4b  & \citet{kepler4b}  &       $0.50^{+0.06}_{-0.06}$ &                       & 1.21 & $0.53^{+0.10}_{-0.08}$ & \\ 
Kepler-5b  & \citet{kepler5b}  &       $0.33^{+0.02}_{-0.02}$ &                       & 1.29 & $0.34^{+0.04}_{-0.03}$ & \\
Kepler-6b  & \citet{kepler6b}  &       $0.63^{+0.03}_{-0.02}$ &                       & 1.35 & $0.63^{+0.05}_{-0.03}$ & \\ 
Kepler-7b  & \citet{kepler7b}  &       $0.30^{+0.03}_{-0.02}$ &                       & 1.15 & $0.30^{+0.04}_{-0.04}$ & \\
Kepler-8b  & \citet{kepler8b}  &       $0.52^{+0.05}_{-0.05}$ &                       & 1.10 & $0.52^{+0.08}_{-0.06}$ & \\
OGLE-TR-10b& \citet{ogle10b}         &                      & $1.08^{+0.11}_{-0.18}$ &      & $1.19^{+0.25}_{-0.22}$ & \\ 
OGLE-TR-111& \citet{ogle111b}        &                      & $1.97^{+0.28}_{-0.26}$ &      & $2.00^{+0.16}_{-0.14}$ & \\
OGLE-TR-113& \citet{ogle113b}        &                      & $2.52^{+0.70}_{-0.12}$ &      & $2.35^{+0.66}_{-0.50}$ & \\ 
OGLE-TR-132& \citet{ogle132b,southw} &                      & $0.71^{+0.22}_{-0.18}$ &      & $0.74^{+0.15}_{-0.12}$ & \\ 
OGLE-TR-182& \citet{ogle182b}        &                      &                      &      & $1.09^{+0.20}_{-0.46}$ & \\ 
OGLE-TR-211& \citet{ogle211b}        &  $0.22^{+0.24}_{-0.05}$ & $0.43^{+0.31}_{-0.05}$ &      & $0.43^{+0.06}_{-0.13}$ & \\
OGLE-TR-56b& \citet{ogle56b,southw}  &                      & $0.886^{+0.31}_{-0.25}$ &    & $1.09^{+0.19}_{-0.16}$ & \\ 
OGLE2-TR-L9& \citet{ogle2L9b}        &                      &                       &    & $0.60^{+0.06}_{-0.05}$ & \\ 
TrES-1     & \citet{tres1}           &  $2.36^{+0.108}_{-0.123}$ & $2.35^{+0.10}_{-0.10}$ & 1.58 & $2.36^{+0.23}_{-0.21}$ & \\
TrES-2     & \citet{tres2}           &                        & $1.38^{+0.07}_{-0.06}$ & 1.40 & $1.38^{+0.17}_{-0.16}$ & 1.3233 $\pm 0.0027$\\ 
TrES-3     & \citet{tres34}          &                        & $2.31^{+0.07}_{-0.06}$ & 1.45 & $2.30^{+0.21}_{-0.17}$ & \\
TrES-4     & \citet{tres34}          &                        & $0.31^{+0.03}_{-0.03}$ & 1.00 & $0.32^{+0.05}_{-0.05}$ & \\
WASP-1b    & \citet{wasp1-2b} &  $0.50^{+0.03}_{-0.04}$ &                       & 1.20 & $0.53^{+0.12}_{-0.05}$ & \\
WASP-2b    & \citet{wasp1-2b} &  $2.00^{+0.20}_{-0.19}$ &                       & 1.67 & $2.08^{+0.48}_{-0.25}$ & \\ 
WASP-3b    & \citet{wasp3b}   &  $0.78^{+0.18}_{-0.07}$ &                       & 0.96 & $0.78^{+0.26}_{-0.11}$ & \\
WASP-4b    & \citet{wasp4-5b} &  $1.81^{+0.01}_{-0.02}$ &                       & 1.50 & $1.80^{+0.29}_{-0.25}$ & \\
WASP-5b    & \citet{wasp4-5b} &  $1.17^{+0.18}_{-0.12}$ &                       & 1.32 & $1.13^{+0.16}_{-0.14}$ & \\
WASP-6b    & \citet{wasp6b}   &  $2.22^{+0.48}_{-0.12}$ &                       & 1.52 & $1.89^{+0.28}_{-0.23}$ & \\
WASP-7b    & \citet{wasp7b}   &  $0.96^{+0.03}_{-0.08}$ &                       & 1.01 & $0.96^{+0.13}_{-0.19}$ & \\ 
WASP-10b   & \citet{wasp10b}  &  $3.32^{+0.10}_{-0.08}$ & $3.13^{+0.10}_{-0.07}$   & 1.85 & $3.11^{+0.24}_{-0.19}$ & \\ 
WASP-11b   & \citet{wasp11b}  &  $2.69^{+0.03}_{-0.17}$ &                       & 1.78 & $2.68^{+0.50}_{-0.48}$ & \\ 
WASP-12b   & \citet{wasp12b}  &  $0.50^{+0.04}_{-0.04}$ &                       & 1.12 & $0.49^{+0.07}_{-0.06}$ & \\
WASP-13b   & \citet{wasp13b}  &  $0.59^{+0.15}_{-0.12}$ &                       & 1.23 & $0.60^{+0.19}_{-0.15}$ & \\ 
WASP-14b   & \citet{wasp14b}  &  $0.76^{+0.08}_{-0.08}$ &                       & 0.98 & $0.77^{+0.17}_{-0.13}$ & \\ 
WASP-15b   & \citet{wasp15b}  &  $0.52^{+0.05}_{-0.04}$ &                       & 1.03 & $0.52^{+0.10}_{-0.09}$ & \\
WASP-16b   & \citet{wasp16b}  &  $1.70^{+0.15}_{-0.21}$ &                       & 1.42 & $1.70^{+0.34}_{-0.35}$ & \\ 
WASP-17b   & \citet{wasp17b}  &  $0.67^{+0.26}_{-0.33}$ &                       & 1.03 & $0.65^{+0.34}_{-0.22}$ & \\
WASP-18b   & \citet{wasp18b}  &  $0.97^{+0.07}_{-0.12}$ &                       & 1.08 & $0.97^{+0.13}_{-0.11}$ & \\ 
WASP-19b   & \citet{wasp19b}  &  $1.61^{+0.11}_{-0.11}$ &                       & 1.50 & $1.63^{+0.27}_{-0.26}$ & \\ 
XO-1b      & \citet{xo1b}     &  $1.76^{+0.04}_{-0.07}$ &                       & 1.46 & $1.77^{+0.09}_{-0.11}$ & \\
XO-2b      & \citet{xo2b}     &                       & $1.49^{+0.11}_{-0.05}$  & 1.51 & $1.47^{+0.09}_{-0.12}$ & \\ 
XO-3b      & \citet{xo3b}     &  $0.63^{+0.05}_{-0.045}$ & $0.66^{+0.09}_{-0.08}$  & 0.90 & $0.66^{+0.14}_{-0.11}$ & \\
XO-4b      & \citet{xo4b}     &                       & $0.51^{+0.04}_{-0.04}$  & 1.03 & $0.49^{+0.05}_{-0.04}$ & \\
XO-5b      & \citet{xo5b}     &  $0.99^{+0.10}_{-0.07}$ & $0.98^{+0.11}_{-0.10}$  & 0.99 & $0.99^{+0.12}_{-0.11}$ & \\
\end{longtable}
\normalsize

\acknowledgments

This publication makes use of data products from the Two Micron All
Sky Survey, which is a joint project of the University of
Massachusetts and the Infrared Processing and Analysis
Center/California Institute of Technology, funded by the National
Aeronautics and Space Administration and the National Science
Foundation. B. Tingley and H.~J. Deeg acknowledge support by grant
ESP2007-65480-C02-02 of the Spanish Ministerio de Ciencia e
Innovaci\'on.

\clearpage


\begin{thebibliography}{}
\bibitem[Aigrain et al.(2008)]{corot4b} Aigrain, S. et al. 2008, \aap,
  488, L43
\bibitem[Alonso et al.(2008)]{corot2b} Alonso, R. et al. 2008, \aap, 482, L21 
\bibitem[Anderson et al.(2010)]{wasp17b} Anderson, D.~R. et al. 2010,
  \apj, 709, 159
\bibitem[Bakos et al.(2004)]{bak2004} Bakos, G.~\'{A}. et al. 2004,
  PASP, 116, 266
\bibitem[Bakos et al.(2007)]{hat5b} Bakos, G. et al. 2007, \apj, 671,
  L173
\bibitem[Bakos et al.(2009a)]{hat10b} Bakos, G.~\'{A}. et al. 2009a,
  \apj, 696, 1950
\bibitem[Bakos et al.(2009b)]{hat13b} Bakos, G.~\'{A}. et al. 2009b,
  \apj, 707, 446
\bibitem[Bakos et al.(2010)]{hat11b} Bakos, G.~\'{A}. et al. 2010,
  \apj, 710, 1724
\bibitem[Barnes (2007)]{bar2007} Barnes, J.~W. 2007, \pasp, 119, 986
\bibitem[Bessel \& Brett (1988)]{bes1988} Bessel, M.~S. \& Brett,
  J.~M., PASP, 100, 1134
\bibitem[Bonomo et al.(2010)]{corot10b} Bonomo, A.~S. et al. 2010,
  \aap, accepted
\bibitem[Barbieri et al.(2009)]{bar2009a} Barbieri, M. et al. 2009,
  \aap, 503, 601
\bibitem[Bord\'e et al.(2010)]{corot8b} Bord\'e, P. et al. 2010, \aap,
  submitted
\bibitem[Borucki et al.(2010)]{kepler4b} Borucki, W.~J. et al. 2010,
  \apj, 713, L126
\bibitem[Bouchy et al.(2010)]{wasp21b} Bouchy, F. et al. 2010, \aap, accepted
\bibitem[Brown (2010)]{bro2010} Brown, T.~M. 2010, \apj, 709, 535
\bibitem[Burke (2008)]{bur2008} Burke, C.~J. 2008, \apj, 679, 1566
\bibitem[Cabrera et al.(2009)]{cab2009} Cabrera, J. et al. 2009, \aap, 506, 501
\bibitem[Cabrera et al.(2010)]{corot13b} Cabrera, J. et al. 2010,
  \aap, submitted
\bibitem[Carpano et al.(2009)]{cap2009} Carpano, S. et al. 2009. \aap, 506, 491
\bibitem[Carter et al.(2009)]{149026b} Carter, J.~C. et al. 2009,
  \apj, 696, 241
\bibitem[Charbonneau et al. (2007)]{wasp1-2b} Charbonneau, D. et
  al. 2007, \apj, 658, 1322
\bibitem[Charbonneau et al.(2009)]{gj1214b} Charbonneau, D. et
  al. 2009, Nature, 462, 891
\bibitem[Christensen-Dalsgaard et al.(2010)]{chr2010}
  Christensen-Dalsgaard, J. et al. 2010, \apj, 713, L164
\bibitem[Cox (2000)]{cox00} Cox, A.~N. 2000, Allen's Astrophysical
  Quantities (New York: Springer)
\bibitem[Deeg et al.(2010)]{corot9b} Deeg, H.~J. et al. 2010, Nature,
  464, 384
\bibitem[Deleuil et al.(2008)]{corot3b} Deleuil, M. et al. 2008, \aap,
  491, 889
\bibitem[Dunham et al.(2010)]{kepler6b} Dunham, E.~W. et al. 2010,
  \apj, 713, L136
\bibitem[Fernandez et al.(2009)]{xo2b} Fernandez, J.~M. et al. 2009,
  \aj, 137, 4911
\bibitem[Ford, Quinn \& Veras (2008)]{for2008} Ford, E.~B., Quinn,
  S.~N., \& Veras, D.  2008, \apj, 678, 1407
\bibitem[Frilund et al.(2010)]{corot6b} Fridlund, M. et al. 2010,
  \aap, 512, 14
\bibitem[Gillon et al.(2007)]{ogle132b} Gillon, M. et al. 2007, \aap, 466, 743
\bibitem[Gillon et al.(2008)]{gil2008a} Gillon, M. et al. 2008, \aap, 485, 871
\bibitem[Gillon et al.(2009a)]{corot1b} Gillon, M. et al. 2009a, \aap,
  506, 359
\bibitem[Gillon et al.(2009b)]{wasp4-5b} Gillon, M. et al. 2009b,
  \aap, 496, 259
\bibitem[Gillon et al.(2009c)]{wasp6b} Gilleon, M. et al. 2009c, \aap, 501, 785
\bibitem[Gillon et al.(2010)]{corot12b} Gillon, M. et al. 2010, \aap, accepted
\bibitem[Gandolfi et al.(2010)]{corot11b} Gandolfi, D. et al. 2010,
  \aap, accepted
\bibitem[Hartman et al.(2009)]{hat12b} Hartman, J.~D. et al. 2009,
  \apj, 706, 785
\bibitem[Hebb et al.(2009)]{wasp12b} Hebb, L. et al. 2009, \apj, 693, 1920
\bibitem[Hebb et al.(2010)]{wasp19b} Hebb, L. et al. 2010, \apj, 708, 224
\bibitem[H\'{e}brard et al.(2010)]{80606b} H\'{e}brard, G. et
  al. 2010, \aap accepted, arXiv:1004.0790
\bibitem[Hellier et al.(2009)]{wasp7b} Hellier, C. et al. 2009, \apj,
  690, L89
\bibitem[Holman et al.(2006)]{xo1b} Holman, M.~J. et al. 2006, \apj, 652, 1715
\bibitem[Holman et al.(2007)]{ogle10b} Holman, M.~J. et al. 2007,
  \apj, 655, 1103
\bibitem[Jenkins et al.(2010)]{kepler8b} Jenkins, J.~M. et al. 2010,
  \apj, submitted
\bibitem[Kipping (2008)]{kip2008} Kipping, D.~M. 2008, \mnras, 389, 1383
\bibitem[Kipping (2010)]{kip2010} Kipping, D.~M. 2010, \mnras, 407, 301
\bibitem[Johnson et al.(2009)]{wasp10b} Johnson, J.~A. et al. 2009,
  \apj, 692, L100
\bibitem[Joshi et al.(2009)]{wasp14b} Joshi, Y.~C. et al. 2009,
  \mnras, 392, 1532
\bibitem[e.g. Kjelsen, Bedding, \& Christensen-Dalsgaard
  (2008)]{kje2008} Kjeldsen, H., Bedding, T.~R., \&
  Christensen-Dalsgaard, J. 2008, \apj, 683, L175
\bibitem[Koch et al.(2010)]{kepler5b} Koch, D.~G. et al. 2010, \apj, 713, L131
\bibitem[Kov\'{a}cs et al.(2007)]{hat4b} Kov\'{a}cs, G. et al. 2007,
  \apj, 670, L41
\bibitem[L\'{e}ger et al.(2009)]{corot7b} L\'{e}ger, A. et al. 2009,
  \aap, 506, 287
\bibitem[Latham et al.(2009)]{hat8b} Latham, D.~W. et al. 2009, \apj,
  704, 1107
\bibitem[Latham et al.(2010)]{kepler7b} Latham, D.~W. et al. 2010,
  \apj, 713, L140
\bibitem[Laughlin et al.(2009)]{lau2009} Laughlin, G. et al. 2009,
  Nature, 457, 562
\bibitem[Lister et al.(2009)]{wasp16b} Lister, T.~A. et al. 2009,
  \apj, 703, 752
\bibitem[Mann, Gaidos, \& Gaudi (2010)]{man2010} Mann, A.~W., Gaidos,
  E., \& Gaudi, S.~B. 2010, \apj, 719, 1454
\bibitem[McCullough et al.(2008)]{xo4b} McCullough, P.~R. et al. 2008,
  \apj, submitted
\bibitem[Noyes et al.(2008)]{hat6b} Noyes, R.~W. et al. 2008, \apj, 673, L79
\bibitem[P\'{a}l et al.(2008)]{pal2008} P\'{a}l, A. et al. 2008 \apj, 680, 1450
\bibitem[P\'{a}l et al.(2009)]{xo5b} P\'{a}l , A. et al. 2009, \apj, 700, 783
\bibitem[P\'{a}l et al.(2010)]{hat2b} P\'{a}l, A. et al. 2010, \mnras,
  401, 2665
\bibitem[Pietrukowicz et al.(2010)]{ogle113b} Pietrukowicz, P. et
  al. 2010, \aap, 509, A4
\bibitem[Pilat-Lohinger (2009)]{pil2009} Pilat-Lohinger, E. 2009, IJAsB, 8, 175
\bibitem[Pollacco et al.(2006)]{poll2006} Pollacco, D. et al. 2008,
  PASP, 118, 1407
\bibitem[Pollacco et al.(2008)]{wasp3b} Pollacco, D. et al. 2008,
  \mnras, 385, 1576
\bibitem[Pont et al.(2005)]{pon2005} Pont, F. et al. 2005, \aap, 438, 1123 
\bibitem[Pont et al.(2008)]{ogle182b} Pont, F. et al. 2008, \aap, 487, 749 
\bibitem[Queloz et al.(2009)]{que2009} Queloz, D. et al. 2009, \aap, 506, 303
\bibitem[Rauer et al.(2009)]{corot5b} Rauer, H. et al. 2009, \aap, 506, 281
\bibitem[Sackett (1999)]{sac1999} Sackett, P.~D. 1999, in Planets
  Outside the Solar System: Theory and Observations,
  ed. J. M. Mariotti \& D. Alloin (NATO ASI Ser. C, 532; Dordrecht:
  Kluwer), 189
\bibitem[Sirko \& Pacy\'{n}ski (2003)]{sir2003} Sirko, E. \&
  Paczy\'{n}ski, B. 2003, \apj, 592, 1217
\bibitem[Santos et al.(2006)]{ogle56b} Santos, N.~C. et al. 2006,
  \aap, 450, 825
\bibitem[Seager \& Mall\'{e}n-Ornelas (2003)]{sea2003} Seager, S. \&
  Mall\'{e}n-Ornelas, G. 2003, \apj, 585, 1038
\bibitem[Shporer et al.(2009a)]{gj436b} Shporer, A. et al. 2009a,
  \apj, 694, 1559
\bibitem[Shporer et al.(2009b)]{hat9b} Shporer, A. et al. 2009b, \apj,
  690, 1393
\bibitem[Skillen et al.(2009)]{wasp13b} Skillen, I. et al. \aap, 502, 391
\bibitem[Skrutskie et al.(2006)]{skr06} Skrutskie, M.~F. et al. 2006,
  \aj, 131, 1163
\bibitem[Smalley et al.(2010)]{wasp26b} Smalley, B. et al. 2010, \aap,
  submitted
\bibitem[Snellen et al.(2009)]{ogle2L9b} Snellen, I.~A.~G. et
  al. 2009, \aap, 497, 545
\bibitem[Southworth (2008)]{southw} Southworth, J. 2008, \mnras, 386, 1644  
\bibitem[Southworth et al.(2009)]{wasp18b} Southworth, J. et al. 2009,
  \apj, 707, 167
\bibitem[Sozzetti et al.(2007)]{tres2} Sozzetti, A. et al. 2007, \apj,
  664, 1190
\bibitem[Sozzetti et al.(2009)]{tres34} Sozzetti, A. et al. 2009,
  \apj, 691 1145
\bibitem[Tassoul (1980)]{tas1980} Tassoul, M. 1980, \apjs, 43, 469
\bibitem[Tingley \& Sackett (2005)]{tin2005} Tingley, B. \& Sackett,
  P.~D. 2005, \apj, 627, 1011
\bibitem[Tingley et al.(2010)]{corot14b} Tingley, B. et al. 2010,
  \aap, submitted
\bibitem[Torres et al.(2007)]{hat3b} Torres, G. et al. 2009, \apj, 666, L121 
\bibitem[Udalski et al.(2002a)]{uda2002a} Udalski, A. et al. 2002, AcA, 52, 1
\bibitem[Udalski et al.(2002b)]{uda2002b} Udalski, A. et al. 2002, AcA, 52, 115
\bibitem[Udalski et al.(2002c)]{uda2002c} Udalski, A. et al. 2002, AcA, 52, 317
\bibitem[Udalski et al.(2003)]{uda2003} Udalski, A. et al. 2003, AcA, 53, 133
\bibitem[Udalski et al.(2008)]{ogle211b} Udalski, A. et al. 2008,
  \aap, 482, 299
\bibitem[Vandakurov (1967)]{van1967} Vandakurov, Yu.~V. 1967, AZh, 44,
  786 (Eng. Transl.: Sov. Astron., 11, 630)
\bibitem[Weldrake et al.(2008)]{lupus3b} Weldrake, D.~T.~W. et
  al. 2008, \apj, 675, 37
\bibitem[West et al.(2009a)]{wasp11b} West, R.~G. et al. 2009, \aap, 502, 395
\bibitem[West et al.(2009b)]{wasp15b} West, R.~G. et al. 2009, \aj, 137, 4834
\bibitem[Winn et al.(2007a)]{189773b} Winn, J.~N. et al. 2007a, \aj,
  133, 1828
\bibitem[Winn et al.(2007b)]{hat1b} Winn, J. et al. 2007b, \apj, 134, 1707
\bibitem[Winn et al.(2008)]{xo3b} Winn, J.~N. et al. 2008, \apj, 683, 1076
\bibitem[Winn et al.(2009a)]{win2009a} Winn, J. et al. 2009a, \apj, 703, L99
\bibitem[Winn et al.(2009b)]{win2009b} Winn, J. et al. 2009b, \apj, 703, 2091
\bibitem[Winn et al.(2009c)]{win2009c} Winn, J. et al. 2009c, \apj, 693, 794
\bibitem[Winn et al.(2009e)]{17156b} Winn, J.~N. et al. 2009e, \apj,
  693, 794
\bibitem[Winn, Holman \& Fuentes (2007)]{ogle111b} Winn, J.~N.,
  Holman, M.~J. \& Fuentes, C.~I. 2007, \aj, 133, 11
\bibitem[Winn, Holman \& Roussanova (2007)]{tres1} Winn, J.~N.,
  Holman, M.~J. \& Roussanova, A. 2007, \apj, 657, 1098
\bibitem[Yee \& Gaudi (2008)]{yee2008} Yee, J.~C. \& Gaudi, S.~B. 2008,
  \apj, 688, 616

\end{thebibliography}
\end{document}